\documentclass[aoas,MSNbibl,nameyear,dvips]{arximspdf}

\doi{10.1214/14-AOAS728A} 
\volume{8}
\issue{4}
\pubyear{2014}
\firstpage{1947}
\lastpage{1951}
\docsubty{FLA}
\referstodoi{10.1214/14-AOAS728}


\begin{document}
\begin{frontmatter}

\title{Discussion of ``Spatial accessibility of pediatric primary
healthcare: Measurement and inference''}
\runtitle{Discussion}

\begin{aug}
\author[A]{\fnms{Laura A.}~\snm{Hatfield}\corref{}\ead[label=e1]{hatfield@hcp.med.harvard.edu}}
\runauthor{L.~A. Hatfield}
\affiliation{Harvard Medical School}
\address[A]{Department of Health Care Policy\\
Harvard Medical School\\
180 Longwood Ave\\
Boston, Massachusetts 02115\\
USA\\
\printead{e1}}
\end{aug}

\received{\smonth{7} \syear{2014}}


\end{frontmatter}
The Affordable Care Act's Medicaid expansions aim to provide primary care
to more Americans by changing eligibility criteria and payments to
providers. While these reduce financial barriers, Medicaid patients are
still more likely than privately insured to report other barriers,
including lack of transportation, lack of timely appointments, long clinic
waiting times and limited hours [\citet{Cheetal12}]. And removing
financial barriers does not necessarily increase primary care utilization;
in fact, following health care reform in Massachusetts, emergency
department care actually increased [\citet{PBYanLan14}]. Other authors
document preferences for emergency department care among low-income
individuals [\citet{Kanetal13}]. Most relevant to this discussion is
that patients reported that hospitals were more accessible than ambulatory
primary care.

In the paper under discussion, \citet{NobSerSwa14} focus on
\textit{spatial accessibility} of primary care by developing a
sophisticated method for assigning patients to nearby primary care
physicians, studying three measures of accessibility given these
assignments, and fitting spatially varying coefficient models to understand
how accessibility varies with measurable factors. I will begin by
describing the strengths of the approach and then discuss a few limitations
and extensions.

The authors have gone to considerable trouble to build an assignment model
that accounts for realistic features of both demand and supply sides. Their
approach minimizes travel time subject to realistic constraints on both
sides: physicians require a minimum panel size to stay in business, the
proportion of physicians that accept Medicaid and Medicaid caseload vary,
patients distribute among nearby physicians to minimize congestion, and
patients with (without) cars are willing to travel no more than 10 (25)
miles for primary care. Conditional on the assignments generated by this
procedure, the authors study variation in three access measures:
congestion, coverage (having a physician within the allowed travel
maximum), and travel time by census tract and population (Medicaid vs.
non-Medicaid). They also study policy interventions that alter three key
parameters: the proportion of physicians accepting Medicaid, the proportion
of Medicaid patients in physicians' panels, and the mobility of Medicaid
patients. Finally, the authors consider how their accessibility measures
co-vary across space with factors such as household income, racial
diversity, unemployment and education.

The authors assemble data about physicians (from Centers for Medicaid and
Medicare Services) and the Medicaid population (from the Census Bureau) in
Georgia and apply their assignment method at the census tract level.
Limitations inherent in these data require several simplifications. For
example, the distance from the census tract centroid to physicians' offices
are used as the travel costs for all patients in a given census tract.
Further, the authors lack information on which \textit{particular}
physicians accept Medicaid and their Medicaid caseloads, so they do some
basic calculations using county-level Medicaid acceptance rates and
Medicaid caseload by practice site (public hospital, community health
clinic and private offices). Geographic misalignment limits identifying the
Medicaid-qualified population with access to vehicles. Although the
household income thresholds (by age) can be applied at the census-tract
level, these data do not include information about vehicles. For that, the
authors turn to the American Community Survey data on a 5\% subsample of
households in 63 regions of the state. The authors figure the vehicle rates
for Medicaid-qualified households and other households. Presumably, they
assume that these rates are constant over the census tracts within those
regions. This is a reasonable approach and properly incorporates the
correlation between low income and vehicle ownership.

The authors emphasize that the model is insensitive to the values of
several parameters (Figures~2--5 in Appendix C). Travel cost and congestion
were mostly flat across the ranges of the parameters, or could only get
worse compared to the status quo. Congestion had some room for improvement
with increasing minimum panel size. Moreover, none of the parameters
differentially affected Medicaid patients' access. Differentially affecting
Medicaid patients (in a positive way) is the policy goal, so clearly these
parameters are not fruitful intervention targets. Thus, the authors
examined three further parameters more likely to provide targeted impact:
the proportion of physicians accepting Medicaid, their Medicaid caseloads,
and the mobility of Medicaid patients relative to the rest of the
population. Again, they found very little positive impact on Medicaid
patients, though they were able to make things worse. For instance,
reducing the proportion of physicians that accept Medicaid substantially
increased travel times and reduced coverage.

Surprisingly, decreasing the maximum proportion of Medicaid patients in
physicians' panels across the board \textit{decreased} the congestion
Medicaid patients (who manage to find a physician) face. This is not
surprising given the correlation (described below) among the three
measures. In general, coverage and travel cost appear to be negatively
correlated with one another, while congestion is negatively related to
coverage and mostly independent of travel cost. An intervention that
increases coverage is likely to be accompanied by increased congestion and
vice versa.

These are important and policy-relevant results because on their face, the
proportion of physicians accepting Medicaid and Medicaid caseloads seem
like reasonable places to look for policy interventions. In particular,
raising Medicaid payment rates is meant to affect both. But as these
authors demonstrate, there may be unintended consequences.

A second important result of the paper is the close correlation of the
three accessibility measures: coverage, congestion, and travel time. In the
three-dimensional space defined by coverage, travel time and congestion,
only some quadrants are represented in the Georgia data. In particular,
they almost never see high travel cost combined with either high coverage
or low congestion. That is, where patients must travel long distances to
find a physician, many do not have a physician at all and if they do, the
congestion is high. Also, they almost never see low travel cost, low
congestion and low coverage. That is, where patients have a nearby
physician and there is little crowding, nearly everyone \textit{has} a
physician.

The maps make it clear that these access issues are strongly associated
with urban/rural location. This is especially true for Medicaid patients.
Either they live in a city with a robust supply of physicians and where
high demand implies many will accept Medicaid, or they live in a rural area
where there are simply too few physicians for the population. As the
authors point out, ``Among the 159 counties in Georgia, approximately $1/3$
have no pediatrician.'' The remaining approximately 107 counties have some
768 pediatricians; these are highly concentrated in the population centers,
as shown by Appendix Figure~10 highlighting census tracts with higher than
the statewide average accessibility. The cities stick out prominently.

The paper's Figure~3 highlights tracts where accessibility among Medicaid
and other patient populations differ significantly. On coverage and travel
costs, the non-Medicaid patients always do better. Counterintuitively, the
Medicaid population has significantly lower congestion than others,
especially in urban centers that are otherwise more congested than the
statewide average. Looking closely, we see a hole in this map in
north-central Atlanta. Examining several public data sources, we find that
this area corresponds to the wealthiest, most educated section of the city
(encompassing the authors' own institution, Georgia Institute of
Technology) and where relatively fewer children live [Carnathan (\citeyear{Car12})]. I
was intrigued to see that the hole fills in and the areas of advantage to
Medicaid patients expands when the assignment model assumes that
Medicaid-accepting physicians take only \textit{half as large a Medicaid
caseload} (Appendix Figure~9). These results are likely driven by inelastic
supply and the limitations on capacity for Medicaid patients. In rural
areas with limited physician supply, the whole population faces long travel
time and low coverage, so reducing Medicaid caseloads frees up coverage
(and reduces travel time) for the non-Medicaid population.

In a physician-rich area such as the city center, patients are
unconstrained by travel costs and can choose the closest from among a
relatively large set of physicians. Very few Medicaid patients are assigned
to physicians in the wealthiest area of town and, therefore, we see no
statistical evidence in those areas for an advantage to Medicaid patients.
When all physicians, including those who practice in poor areas of the
city, accept a lower Medicaid caseload, Medicaid patients may need to
travel to a physician in these wealthier areas of the city where they then
enjoy the same congestion advantages as they do in other areas. In general,
where physician supply is abundant, reducing Medicaid caseloads may require
Medicaid patients to travel slightly further, but when they do, they
experience lower congestion.

The third major contribution of the paper (after the assignment algorithm
and the studies of geographic and population variation in the three
accessibility measures) is a set of spatially varying coefficient models
that attempt to associate accessibility with measurable characteristics:
household income, higher education, unemployment, nonwhite population,
population density, distance to hospitals, and a diversity ratio. The model
fitting is complex, with penalized splines and backfitting, such that it is
not entirely clear what models generated the coefficients in the paper's
Table~3. Also, the scales of the variables are not comparable, so it is
difficult to compare the scale of the effects (or widths of the confidence
intervals). Nevertheless, the directions of the effects are mostly
intuitive. Travel time is lower in areas with higher education, lower
unemployment, more hospital access and more diversity. One exception to the
generally sensible coefficients is that of income, which is
\textit{positively} associated with travel time. Another exception is the
space-varying coefficient for distance to hospitals; the effect actually
changes sign from negative in Atlanta to positive in the most rural areas.
Neither of these results is explained much by the authors. In the real
world, travel cost may be discounted in higher income households, that is,
wealthier families can afford to live farther from physicians. However, the
authors' assignment model does not allow physician choice to depend on
household income, so this cannot explain the results. The authors do
suggest that some fit statistics indicated models were better without
income.

The sign reversal for the distance to hospitals coefficient in Atlanta
versus the rural areas may also be an issue of competition between effects.
In dense urban areas where the distance to hospitals is low for nearly
everyone, models with several other predictors may have trouble allocating
the remaining variation in distance to hospitals in a sensible way.

Population density is the closest covariate to indicating urban/rural and
it is varying in space though negative everywhere,\vadjust{\goodbreak} as we expect (travel
cost increases with decreasing density). The strongest coefficients are in
the most rural areas and nearly zero in the urban centers. This mirrors the
pattern for racial diversity. Both of these indicate that the effects on
travel time are strongest where physician supply is lowest.

There are several interesting extensions and applications of these methods.
The results here suggest that most potential improvement in Medicaid
patients' access to pediatricians is by re-allocating the supply of
physicians to areas where they are needed more. The authors mention
interventions along these lines in Section~4.3, such as loan repayment for
practicing in rural areas and telemedicine. Rather than conditioning on the
locations of physicians and altering their characteristic, these methods
can be used to alter the \textit{locations} of the physicians and apply
allocation methods conditional on various arrangements to study
accessibility.

Another potential application is comparing these optimization approaches to
others used by managed care organizations to assign enrollees to providers
when they do not choose a primary care physician themselves. In this case,
patients' exact locations are known and the supply of physicians is more
limited compared to the Georgia example of the paper. Thus, it could be
useful to compare the current approaches of managed care plans to assigning
new enrollees (those who do not elect a physician) to population-oriented
approaches such as the constrained optimization developed in this paper.
Another related area of application would be to examine the impacts on
access of narrow network policies being implemented by many payers.



%

\printaddresses
\end{document}